%% file: main.tex
\documentclass[conference,letterpaper,10pt]{IEEEtran}
\usepackage[T1]{fontenc}
\usepackage{amsmath,amssymb,bm,cite,url,graphicx,booktabs,array}
\usepackage{placeins}
\usepackage[hidelinks]{hyperref}
\hypersetup{
 pdftitle={Dynamic Channel Gain Map Reconstruction via Object-Attached Gaussian Propagation},
 pdfauthor={Yi Wang; Junting Chen},
 pdfsubject={IEEE ICC 2027 review manuscript},
 pdfkeywords={Channel gain map reconstruction, Gaussian propagation, vehicular communications}
}
\newcommand{\vect}[1]{\boldsymbol{#1}}
\newcommand{\spplus}{\operatorname{softplus}}
\newcommand{\diag}{\operatorname{diag}}
\newcommand{\erf}{\operatorname{erf}}
\begin{document}
\title{Dynamic Channel Gain Map Reconstruction via Object-Attached Gaussian Propagation}
\author{\IEEEauthorblockN{Yi Wang and Junting Chen}
\IEEEauthorblockA{The Chinese University of Hong Kong, Shenzhen, China\\
Email: wangyi2@cuhk.edu.cn, juntingc@cuhk.edu.cn}}
\maketitle
\begin{abstract}
Channel gain maps describe location-dependent propagation, but moving vehicles can change reception beyond their physical footprints. Reconstructing these changes requires connecting local object states to regional power. We consider a fixed transmitter and predefined receiver areas, with historical static and dynamic radio supervision and continuing delayed, noisy vehicle-state messages. We propose OA-GS, an object-attached Gaussian propagation model with a shared area correction. Each vehicle carries anisotropic kernels whose effective attributes are supplied by offline-calibrated templates and affine condition generators. Closed-form Gaussian integrals along finite propagation segments evaluate an effective attenuation model, while directional scattering terms account for additional power. These interactions modulate a learned static reference and contribute to regional channel gain. A shared multilayer perceptron then uses object, propagation, and receiver-area features to predict a bounded path-loss residual that compensates for errors in the compact propagation model. All learned parameters remain fixed during deployment, and reconstruction uses arrived vehicle messages without current-traffic radio measurements. In a fixed-scene ray-tracing study over four development traffic sequences and three initializations, OA-GS reduces map and strong-event dynamic-response root-mean-square errors (RMSEs) from 2.503 to 1.272 dB and from 5.965 to 3.563 dB, respectively, relative to a task-adapted GS-CG baseline receiving eight initial radio reports. The shared correction reduces strong-event dynamic-response error from 3.962 to 3.563 dB under the same propagation backbone, demonstrating the benefit of combining explicit propagation structure with learned regional correction.
\end{abstract}
\input{sections/introduction}
\input{sections/system}
\input{sections/method}

\input{sections/experiments}
\FloatBarrier
\section{Conclusion}
In the evaluated vehicular scene, OA-GS reconstructs channel gain maps from delayed, noisy vehicle messages without current-traffic RF measurements. It combines offline-calibrated Gaussian propagation with shared regional correction to capture reception changes beyond vehicle footprints. Compared with task-adapted GS-CG and MLP-GS, it improves both map accuracy and strong-event dynamic-response prediction. Future work will examine transfer to new environments, moving transmitters and receivers, and validation with measured radio data.

\input{sections/references}
\end{document}

%% file: sections/introduction.tex
\section{Introduction}
Channel gain maps provide location-dependent propagation information for coverage management and link selection \cite{zeng2021,ckmtutorial}. Buildings and terrain establish much of the propagation background, while moving vehicles can block strong components, release earlier obstructions, or add scattered power. Measurements in vehicular networks confirm that intervening vehicles can substantially attenuate reception \cite{boban2011}. These effects can reach receivers outside a vehicle's immediate neighborhood. Dynamic channel gain map reconstruction must therefore reproduce localized changes while retaining accuracy in stable regions; a small average map error alone can conceal a substantial blockage event.

On roads without continuous power monitoring, vehicle-state messages may remain available as traffic changes. Cooperative messages supply position and motion \cite{cam}, whereas regional radio measurements require suitable observation locations and reporting \cite{mdt}. Historical static and traffic-dependent observations can support offline learning, but current vehicle positions alone do not determine effective attenuation or directional scattering. The problem is to reuse this historical knowledge at changing object configurations and reconstruct regional channel gain without acquiring radio-frequency (RF) reports from the current traffic sequence.

RadioUNet predicts path-loss maps from environment descriptions \cite{radiounet}. Three-dimensional Gaussian representations provide learnable geometry and directional attributes \cite{kerbl2023}. WRF-GS and WRF-GS+ combine Gaussian primitives with neural RF representations for spatial-spectrum reconstruction \cite{wrfgs}. PropSplat learns Gaussian path-loss offsets relative to a baseline propagation model from RF measurements without an external scene map \cite{propsplat}. GS-CG represents regional gain through distance attenuation, transmittance, and scattering; its incremental update freezes the static reference and fits residual-initialized active Gaussians to new RF measurements \cite{gscg}. MLP-GS is its lower-complexity neural variant \cite{gscg}. We study regional map reconstruction driven by continuing vehicle messages without current-traffic RF fitting.

Dynamic visual Gaussians organize geometry over time \cite{wu2024,yang2024}. RFScape learns object-level RF interactions and composes them within a ray-tracing pipeline \cite{rfscape}. Our approach evaluates compact representative propagation segments and a regional correction. This requires connecting moving objects to distant receiver areas while compensating for approximation errors without excessive responses in stable regions.

We propose object-attached Gaussian propagation (OA-GS), a propagation model with a shared geometry-conditioned area correction. Historical supervision supplies reusable object templates and affine generators of effective wireless attributes. Vehicle messages transport the kernels; finite-segment interactions modulate retained environment contributions and add directional power. Closed-form integrals evaluate geometric interactions, while a shared neural network corrects remaining area-level errors.

The contributions are summarized as follows:
\begin{itemize}
\item We develop an object-attached propagation representation that connects local vehicle support to distant receiver regions through finite-segment attenuation, directional scattering, and joint static--dynamic power composition.
\item We introduce a bounded geometry-conditioned area correction and quantify its additional contribution through matched comparisons with the same frozen propagation backbone.
\item We evaluate the unified implementation on four development traffic sequences and three initializations, demonstrating improved map and strong-event accuracy over the task-adapted GS-CG and MLP-GS baselines and characterizing regional correction and message-perturbation behavior.
\end{itemize}

%% file: sections/system.tex
\section{System Model and Reconstruction Task}
\subsection{Scenario and Available Information}
We consider a fixed transmitter (Tx), permanent road geometry, moving vehicles, and receiver (Rx) areas $\mathcal A_q$. Offline inputs are public scene/Tx/Rx geometry and historical static and dynamic RF labels paired with vehicle messages. Runtime messages supply identity, type/size, position, heading, and motion. With $\mathcal M_{\leq t}$ denoting arrived messages, we reconstruct the current map without future trajectories or current-sequence RF reports. Exact simulated poses and query labels are evaluation-only.

\subsection{Regional Power and Dynamic Response}
Following the regional-gain formulation in \cite{gscg}, define the regional gain and path loss (PL) for linear gain $G(\vect r,t)$ by
\begin{equation}
\bar G_q(t)=\frac{1}{|\mathcal A_q|}\int_{\mathcal A_q}G(\vect r,t)\,d\vect r,
\quad L_q(t)=-10\log_{10}\bar G_q(t).
\label{eq:target}
\end{equation}
For each realized configuration, \eqref{eq:target} averages linear power spatially. We regress PL under uncertain messages, with state averaging specified in Section~III-B. PL is negative channel gain in dB, giving identical error magnitudes. Relative to initial traffic at $t_0$, define
\begin{equation}
\Delta L_q(t)=L_q(t)-L_q(t_0),\quad
\Delta\hat L_q(t)=\hat L_q(t)-\hat L_q(t_0).
\label{eq:delta}
\end{equation}
Each estimator subtracts its own initial prediction. Positive changes indicate weaker reception. The initial traffic contains vehicles and differs from the historical empty-scene reference.

%% file: sections/method.tex
\section{Object-Attached Gaussian Propagation}
\begin{figure*}[t]
\centering
\includegraphics[width=\textwidth]{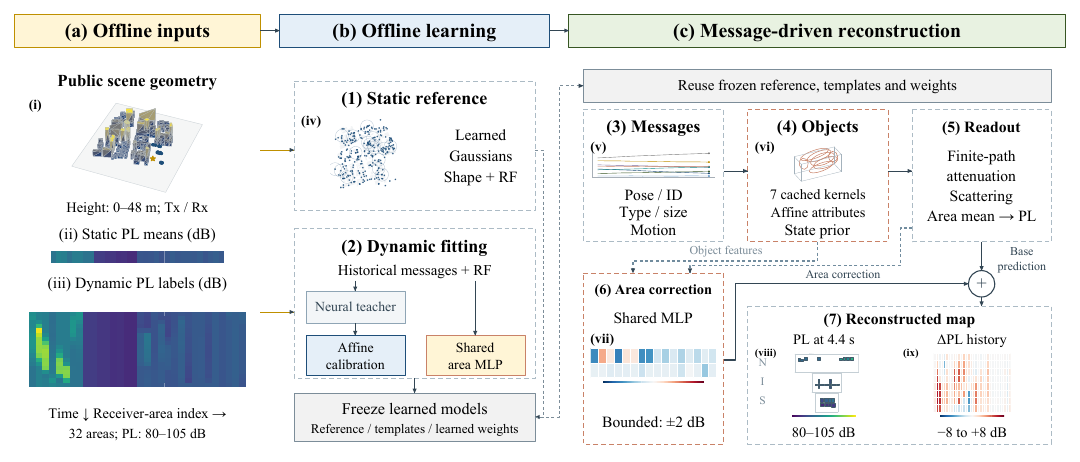}
\caption{OA-GS pipeline: offline supervision calibrates the reference and response models; arrived messages drive propagation and shared area correction. Insets show simulated supervision, learned representations, and path-loss predictions; N/I/S denote receiver windows. Solid arrows indicate prediction or calibration; dashed arrows indicate frozen models or feature context.}
\label{fig:framework}
\end{figure*}

In Fig.~\ref{fig:framework}, affine attribute generation and analytic propagation form the backbone; a shared network corrects its regional output.

\subsection{Static Reference and Object Representation}
The static reference follows GS-CG \cite{gscg}: 256 building-surface and 16 ground seeds are initialized from public geometry, then centers, scales, rotations, opacity, and RF attributes are fitted to historical empty-scene area means and frozen. Its direct and static-Gaussian components use representative Tx--Rx and Tx--Gaussian-center--Rx segments, respectively, connecting local vehicle interactions to distant receivers.

Each vehicle carries seven local kernel roles, one central and six directional. Let $k$ index all vehicle kernels and $o(k)$ its owner. With message-derived object position $\vect p_o$ and orientation $\vect Q_o$, local offset $\vect b_k$, scale $\vect s_k$, and rotation $\vect R_k$ give
\begin{align}
\vect\mu_k&=\vect p_{o(k)}+\vect Q_{o(k)}\vect b_k,\nonumber\\
\vect\Sigma_k&=\vect Q_{o(k)}\vect R_k\diag(\vect s_k^2)\vect R_k^T\vect Q_{o(k)}^T.
\label{eq:pose}
\end{align}
Here $\vect\Sigma_k$ is spatial kernel covariance. Each kernel also carries extinction $\kappa_k\geq0$, scattering strength $\sigma_k\geq0$, symmetric direction matrix $\vect A_k\in\mathbb R^{4\times4}$, and eight segment coefficients $\vect c_k$.

For vehicle type $c$ and kernel role $a$, the conditional response has the affine form
\begin{equation}
\vect\zeta_{ca}(\vect x)=\vect C_{ca}^{T}[1;\tilde{\vect x}],
\label{eq:affine}
\end{equation}
where the nine entries of $\vect x$ comprise body-frame Tx direction, $\log(d/\mathrm m)/5$ with $d\geq0.1$ m, body-frame velocity, yaw rate, and message age. Direction/distance use the nominal type/role anchor. Historical means and standard deviations (floored at 0.05) give $\tilde{\vect x}$; statistics and coefficients remain fixed while generated attributes vary with the configuration.

We combine raw extinction/scattering coordinates and ten unique directional-matrix entries as
\begin{equation}
\vect\eta_k=\vect\eta^{\rm tpl}_{ca}
+\tfrac14\vect\zeta_{ca}(\vect x)
+\vect H_{ca}\vect z_o,
\label{eq:attributes}
\end{equation}
where $\vect\eta^{\rm tpl}_{ca}$ is an offline template and $\vect z_o\in\mathbb R^4$ the wireless state. Cached $\vect H_{ca}$ maps its coordinates to extinction, scattering, zero-mean directional-role extinction contrast, and a directional-matrix mode. With footprint $s_o^{\rm fp}$,
\begin{align}
\kappa_k&=2s_o^{\rm fp}\spplus(\eta_{k,1}-1),\nonumber\\
\sigma_k&=s_o^{\rm fp}\spplus(\eta_{k,2}-2),\qquad
\vect A_k=\operatorname{Sym}_{10}(\vect\eta_{k,3:12}),
\label{eq:constraints}
\end{align}
where $\operatorname{Sym}_{10}$ fills a symmetric $4\times4$ matrix from its ten unique entries. Both $\kappa_k$ and $\sigma_k$ have units of area. A separate affine function generates $\vect c_k=\tfrac14\tanh(\vect D_{ca}^{T}[1;\tilde{\vect x}])\in\mathbb R^8$ for segment-dependent modulation.

Offsets, scales, rotations, and mode weights are cached by type/role and independent of query. Messages and sampled pose perturbations move kernels through \eqref{eq:pose}. Affine generators and deterministic constraints provide effective RF attributes learned from historical supervision, rather than measured material constants; the area network corrects remaining regional error.

\subsection{Analytic Interaction and Regional Readout}
The normalized spatial density of kernel $k$ is
\begin{equation}
\rho_k(\vect r)=\frac{\exp[-\tfrac12(\vect r-\vect\mu_k)^T\vect\Sigma_k^{-1}(\vect r-\vect\mu_k)]}{(2\pi)^{3/2}|\vect\Sigma_k|^{1/2}}.
\label{eq:density}
\end{equation}
For a segment from $\vect x_0$ to $\vect x_1$, let $\ell=\|\vect x_1-\vect x_0\|$ and $\vect e=(\vect x_1-\vect x_0)/\ell$. The accumulated interaction is
\begin{equation}
I_k(\vect x_0,\vect x_1)=\int_0^\ell\rho_k(\vect x_0+\eta\vect e)\,d\eta.
\label{eq:integral}
\end{equation}
Writing $\alpha=\vect e^T\vect\Sigma_k^{-1}\vect e$, $\beta=\vect e^T\vect\Sigma_k^{-1}(\vect x_0-\vect\mu_k)$, $\gamma=(\vect x_0-\vect\mu_k)^T\vect\Sigma_k^{-1}(\vect x_0-\vect\mu_k)$, and $Z_k=(2\pi)^{3/2}|\vect\Sigma_k|^{1/2}$ yields
\begin{equation}
\begin{split}
I_k={}&\frac{e^{-(\gamma-\beta^2/\alpha)/2}}{Z_k}\sqrt{\frac{\pi}{2\alpha}}\left[
\erf\!\left(\frac{\alpha\ell+\beta}{\sqrt{2\alpha}}\right)\right.\\[-1mm]
&\hspace{35mm}\left.-\erf\!\left(\frac{\beta}{\sqrt{2\alpha}}\right)\right].
\end{split}
\label{eq:closed}
\end{equation}
Zero-length segments have zero interaction. Let $\widetilde{\vect R}_k=\vect Q_{o(k)}\vect R_k$ and $\widetilde{\vect e}_k=\widetilde{\vect R}_k^T\vect e$. The closest point in the kernel's Mahalanobis metric is obtained by clipping $-\beta/\alpha$ to $[0,\ell]$. Its offset from the kernel center, expressed in kernel axes and divided by $\vect s_k$, is $\vect v_k^*$. The eight-dimensional segment descriptor is
\begin{equation}
\vect\phi_k=\left[\widetilde{\vect e}_k;
\tfrac13\operatorname{asinh}(\vect v_k^*);
\tfrac15\log(1+\ell/\mathrm m);
\log(1+100\,\mathrm m^2 I_k)\right].
\label{eq:segment_features}
\end{equation}
A bounded factor $g_k=\exp[\tfrac14\tanh(\vect c_k^T\vect\phi_k)]$ conditions the interaction on direction, clearance, length, and overlap. Dynamic power transmittance is
\begin{equation}
T(\vect x_0,\vect x_1)=\exp\!\left[-\sum_k\kappa_k I_k(\vect x_0,\vect x_1)g_k\right].
\label{eq:transmit}
\end{equation}
The units of $\rho_k$, $I_k$, and $\kappa_k$ are $\mathrm{m}^{-3}$, $\mathrm{m}^{-2}$, and $\mathrm{m}^{2}$, respectively. Equation~\eqref{eq:closed} is exact for the specified Gaussian density and segment; the exponential mapping in \eqref{eq:transmit} is an effective attenuation law. GS-CG instead evaluates nearest-point Gaussian overlap on a finite segment \cite{gscg}. Our integral explicitly retains both segment extent and endpoints.

For body-frame directions $\vect u$ and $\vect v$ from a kernel toward Tx and Rx, let $\vect y(\vect u)=[1;\vect u]$ and
$\Psi_k(\vect u,\vect v)=\spplus(\vect y(\vect u)^T\vect A_k\vect y(\vect v))/N_k$, with $N_k$ the positive mean over pairs of six axis directions. The added scattering contribution to channel gain is
\begin{equation}
C_k(\vect r,t)=\frac{\lambda^2\sigma_k\Psi_k(\vect u,\vect v)W_kE_k}{(4\pi)^3d_{\rm in}^2d_{\rm out}^2}.
\label{eq:scatter}
\end{equation}
Here $\lambda$ is wavelength and $d_{\rm in/out}$ are Tx--kernel and kernel--Rx distances, lower-bounded by 0.1 m. For a directional role with body-frame normal $\vect n_k$, define $h_k=4[\vect n_k^T\vect u]_+[\vect n_k^T\vect v]_+$; the central role has $h_k=1$. With $[a]_+=\max(a,0)$, the directional factor is
\begin{equation}
W_k=h_k\exp\!\left[\tfrac14\tanh\!\left(
\vect c_{k,1:3}^T\vect u+\vect c_{k,4:6}^T\vect v\right)\right].
\label{eq:scatter_gate}
\end{equation}
$N_k$ averages the numerator over the 36 pairs from $\{\pm\vect e_1,\pm\vect e_2,\pm\vect e_3\}$. Static power transmittance gives
\[
E_k=T_{\rm stat}^{\rm P}(\mathrm{Tx},\vect\mu_k)
T_{\rm stat}^{\rm P}(\vect\mu_k,\vect r).
\]
The static field-amplitude transmittance $\tau_{\rm stat}$ is squared once: $T_{\rm stat}^{\rm P}(\vect a,\vect b)=|\tau_{\rm stat}(\vect a,\vect b)|^2$. Dynamic $T$ in \eqref{eq:transmit} already acts on power. Separate $\kappa_k$ and $\sigma_k$ control extinction and scattering; $\vect A_k$ couples incoming/outgoing directions, complementing GS-CG's opacity and outgoing-direction representation \cite{gscg}.

\begin{figure}[t]
\centering
\includegraphics[width=\columnwidth]{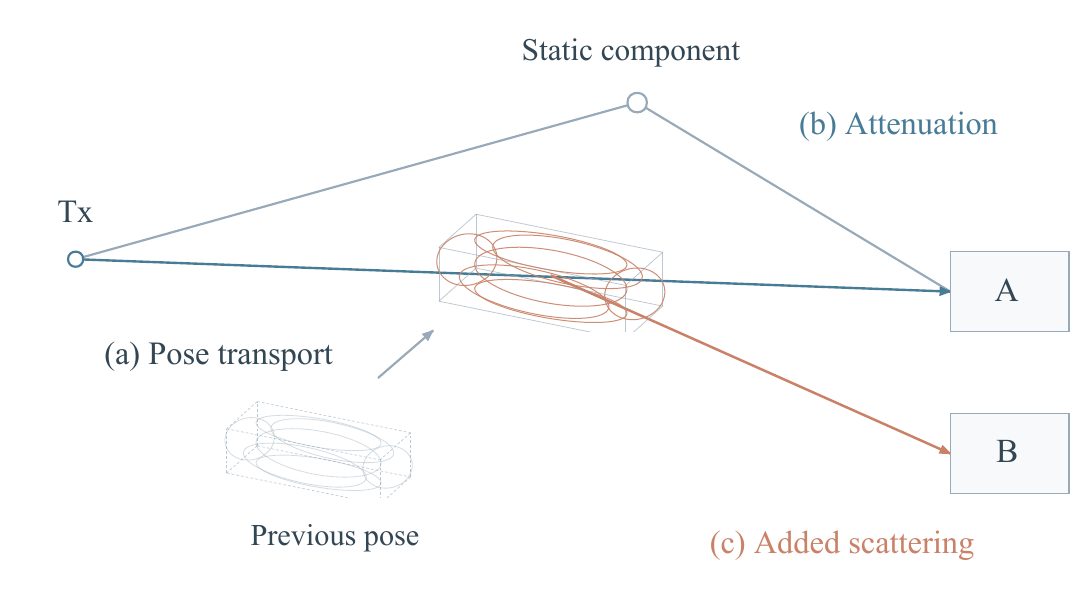}
\caption{Object-attached kernels affect distant receivers: finite-segment attenuation modifies retained power toward A, while directional scattering adds power toward B. Representative segments replace full current-scene ray enumeration.}
\label{fig:mechanism}
\end{figure}

Let $G_0^{\rm ref}(\vect r)$ be the frozen direct gain and $G_j^{\rm ref}(\vect r)$, $j\geq1$, the frozen gain associated with static Gaussian center $\vect\nu_j$. Their dynamic multipliers are
\[
\begin{aligned}
T_0(\vect r,t)&=T(\mathrm{Tx},\vect r),\\
T_j(\vect r,t)&=T(\mathrm{Tx},\vect\nu_j)T(\vect\nu_j,\vect r),\quad j\geq1.
\end{aligned}
\]
The gain prediction is
\begin{equation}
\begin{split}
\hat G(\vect r,t)={}&\sum_jG_j^{\rm ref}(\vect r)T_j(\vect r,t)\\
&+\sum_k C_k(\vect r,t)T_{-o(k)}(\mathrm{Tx},\vect\mu_k)
T_{-o(k)}(\vect\mu_k,\vect r),
\end{split}
\label{eq:compose}
\end{equation}
where $T_{-o(k)}$ excludes the contributing object and the reference sum includes $j=0$. Objects attenuate retained and added power; moving away can restore reception (Fig.~\ref{fig:mechanism}). Representative segments, directional scattering, and noncoherent composition approximate propagation. New diffracted rays and higher-order inter-vehicle scattering are not enumerated; effective attributes and area correction approximate unresolved effects within the training regime.

Let $q_\psi(\vect\xi\mid\mathcal M_{\leq t})$ be the model's joint state prior, specified in Section~III-D. For $N$ samples $\vect\xi_n$ and $M$ spatial points $\vect r_{qm}$ in area $q$, define
\begin{equation}
\begin{aligned}
b_q^{(n)}(t)&=-10\log_{10}\!\left[\frac1M\sum_{m=1}^M\hat G(\vect r_{qm},t;\vect\xi_n)\right],\\
B_q(t)&=\frac1N\sum_{n=1}^N b_q^{(n)}(t)\\
&\approx\mathbb E_{\vect\xi\sim q_\psi(\cdot\mid\mathcal M_{\leq t})}[b_q(t;\vect\xi)],
\end{aligned}
\label{eq:readout}
\end{equation}
where $b_q(t;\vect\xi_n)=b_q^{(n)}(t)$. Spatial averaging forms linear regional gain; state averaging predicts PL under the prior. In general,
\[
\mathbb E[-10\log_{10}\bar G]\ne-10\log_{10}\mathbb E[\bar G].
\]
We use the left-hand quantity for dB regression. Learned supports avoid a complete current-scene ray search and detailed material inputs.

\subsection{Geometry-Conditioned Area Correction}
A shared multilayer perceptron (MLP) corrects regional approximation errors from 48 features: object/kernel context (28), wireless state (4), body-frame Rx direction (3), distance/support/extinction/scattering summaries (4), static context (8), and sample-specific backbone PL $b_q^{(n)}(t)/100$ (1), evaluated before state averaging. Let $\bar{\vect f}_{qok}^{(n)}$ be the spatially averaged features for state $\vect\xi_n$. For $O_t$ vehicles with kernel sets $\mathcal K_o$, define
\begin{align}
a_q(t)&=\frac1N\sum_{n=1}^N\frac1{\sqrt{\max(1,O_t)}}\sum_{o=1}^{O_t}\frac17\sum_{k\in\mathcal K_o}h_\theta(\bar{\vect f}_{qok}^{(n)}),\nonumber\\
\delta_q(t)&=2\tanh a_q(t),\quad
\hat L_q(t)=B_q(t)+\delta_q(t).
\label{eq:residual}
\end{align}
For $O_t=0$, $a_q=\delta_q=0$. The gain $\widehat{\bar G}_q(t)=10^{-\hat L_q(t)/10}$ is the point estimate corresponding to PL, not a conditional mean of linear gain. The $48$--$64$--$64$--$1$ MLP has two sigmoid linear unit (SiLU) activations and 7,361 shared parameters. Spatial features are averaged before the MLP, state scores before tanh. The $\pm2$-dB map correction contributes at most $\pm4$ dB to $\delta_q(t)-\delta_q(t_0)$.

\subsection{Offline Calibration and Causal Deployment}
Offline calibration fits a neural propagation teacher, freezes it to train the shared correction, then initializes affine generators from its historical-input outputs and calibrates them against RF labels. Templates are cached and correction weights reused unchanged. Deployment uses affine generators and constraints in place of the teacher's attribute MLPs; training settings appear in Section~\ref{sec:training}.

At time $t$, the latest two arrived packets provide finite-difference velocity, capped horizontally at 15 m/s, to extrapolate position. Heading uses the latest packet; yaw rate and message age condition the attributes. No future packet is accessed.

Each object's seven-dimensional state contains four wireless coordinates and horizontal-position/yaw perturbations. Stacking these states gives Gaussian prior $q_\psi$ with mean $\vect m_t$ and covariance $\vect P_t$. It starts at zero mean with $\vect P_0=\diag(\vect d_0)$; a type/motion-conditioned affine function generates positive $\vect d_0$. State covariance $\vect P_t$ differs from spatial kernel covariance $\vect\Sigma_k$. No current-sequence RF measurement update or online Bayesian correction is performed. All learned parameters remain frozen.

For each object, the transition factors over $\Delta t$ seconds are
$\vect f=[1,1,1,1,e^{-0.05\Delta t},e^{-0.05\Delta t},e^{-5\Delta t}]^T$.
Stacking them over objects gives $\vect F=\diag(\vect f)$ and
\begin{align}
\vect m_t&=\vect F\vect m_{t^-},\nonumber\\
\vect P_t&=\vect F\vect P_{t^-}\vect F^T+
\diag\!\left(\vect d_0\odot(1-\vect f^2)+\epsilon\Delta t\right),
\label{eq:state}
\end{align}
where $\epsilon=10^{-5}$ per second in state-coordinate units, with a $10^{-7}$ diagonal floor. Initial type/motion generates $\vect d_0$ once; later messages update nominal pose while \eqref{eq:state} propagates uncertainty. Antithetic normal samples with the Cholesky factor of $\vect P_t$ implement \eqref{eq:readout}. These model covariances have not been calibrated to true pose or channel-error distributions.

%% file: sections/experiments.tex
\section{Experimental Evaluation}
\subsection{Scene, Supervision, and Information Protocol}
\begin{figure}[!t]
\centering
\includegraphics[width=\columnwidth]{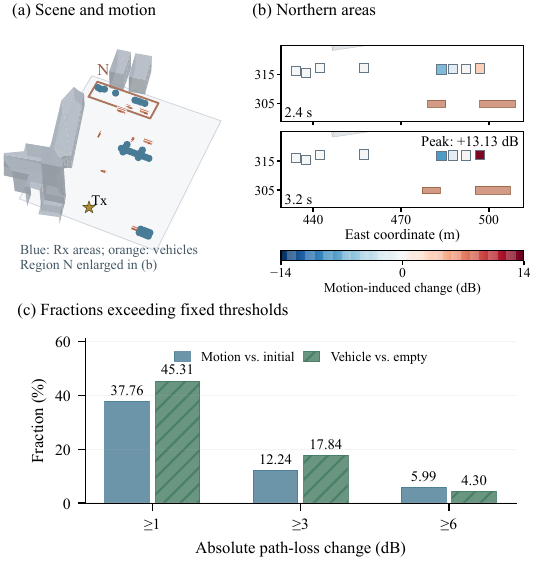}
\caption{Simulated vehicle influence. (a) Scene and eight vehicles. (b) Northern areas at two times; symbols are enlarged without interpolation. (c) Path-loss-change exceedance relative to initial traffic and a matched empty scene: the same 768 future area--time samples, separate from the development performance benchmark.}
\label{fig:impact}
\end{figure}

We use a fixed outdoor scene at 5.9 GHz with Sionna RT and a label pipeline including direct, reflected, diffuse, and diffracted contributions \cite{sionna}. Tx is at $(502.5,142.5,10)$ m. Each of 32 receiver areas contains a $4\times4$ quadrature grid with 1.25-m point spacing. The three vehicle templates are a passenger car ($4.8\times2.0\times1.7$ m), delivery van ($6.2\times2.3\times2.7$ m), and city bus ($12.0\times2.6\times3.3$ m). The model receives their type and dimensions through the external configuration interface. The scene layout is shown in Fig.~\ref{fig:impact}(a).

Historical supervision comprises 32 empty-scene regional means and five traffic sequences with 75 configurations, giving 2,400 dynamic regional labels. Eight message-augmentation views reuse these labels. Four development sequences with 8 or 16 vehicles, denoted D8-1, D16-1, D8-2, and D16-2, support method development and the reported comparison. Three optimization initializations give 12 traffic--initialization pairs. Tx, environment, and receiver regions remain fixed; the 24 query regions were covered by historical supervision.

Messages arrive nominally at 2.5 Hz with 0.4-s delay. Each horizontal coordinate has a sequence-fixed Gaussian bias with 1-m standard deviation and autoregressive jitter with coefficient 0.8 and stationary standard deviation 0.5 m; heading noise has 5-degree standard deviation. Evaluation uses nine times from 1.2 to 8.8 s. OA-GS, its backbone, and the local-response baseline receive no current-sequence RF reports. The task-adapted GS-CG and MLP-GS baselines receive eight initial area-level reports, with 0.5-dB noise standard deviation and a 0.5-dB quantization step, and continuing vehicle messages. This current-report budget excludes historical supervision.

\subsection{Training and Baseline Implementations}
\label{sec:training}
Teacher, correction, and affine calibration use 3,000, 1,200, and 400 steps, respectively, with Adam ($3\times10^{-4}$ learning rate, gradient clipping at 10). Steps cycle through five historical sequences and eight message views, each covering 15 times and 32 areas. Readout uses 16 spatial points and four/eight state samples for training/evaluation. Affine initialization uses regularized least squares on teacher outputs.

\begin{table}[t]
\centering
\caption{Map and dynamic-response reconstruction on four development sequences and three initializations.}
\label{tab:main}
\input{tables/main_results}
\par\smallskip
\begin{minipage}{\columnwidth}\footnotesize
Initial RF counts current-traffic area-level reports, excluding historical supervision. RMSEs average 12 pairwise scores; stable-region false-alarm rates (FPRs) pool counts. Baselines use the task adaptations in Section~IV-B.
\end{minipage}
\end{table}

Let $\mathcal L_{\rm abs,Q}$ and $\mathcal L_{\rm abs,S}$ denote mean squared path-loss errors over the historical query and source areas. Let $\mathcal L_{\Delta,\mathcal V}$ be the mean squared change error over future samples in set $\mathcal V$, where $\mathcal Q$, $\mathcal E$, and $\mathcal C$ denote all query samples, strong events, and stable samples. The common objective is
\begin{equation}
\begin{split}
\mathcal L={}&\mathcal L_{\rm abs,Q}+0.5\mathcal L_{\Delta,\mathcal Q}
+\mathcal L_{\Delta,\mathcal E}+0.25\mathcal L_{\Delta,\mathcal C}\\
&+0.25\mathcal L_{\rm abs,S}+10^{-4}\mathcal R,
\end{split}
\label{eq:loss}
\end{equation}
Here $\mathcal R$ is mean squared trainable-parameter magnitude. Historical changes define event masks; empty masks contribute zero.

The task-adapted GS-CG and MLP-GS baselines \cite{gscg} use 272 static components and 112 active kernels. Initial RF residuals guide active initialization, followed by 1,000 source-fitting steps over active geometry, opacity, and RF attributes or latents. Subsequent motion follows a frozen soft box association for GS-CG and nearest-box association for MLP-GS. MLP-GS also receives 4,000 static and 2,000 historical dynamic training steps using 81 causal input frames. Active density is fixed in these task adaptations; adaptive birth and pruning from the original system are not reproduced. The \emph{local-response baseline} (LOCAL) is a shared object-conditioned network predicting signed regional response from local geometry, motion, state, and an eight-dimensional static summary. It uses the same historical labels and messages as OA-GS, while OA-GS exposes full static propagation components to its operator.

Map root-mean-square error (RMSE) uses all nine evaluation times. Dynamic metrics use the eight future times and each method's own initial map. Strong events satisfy $|\Delta L|\geq3$ dB. Stable samples satisfy $|\Delta L|\leq1$ dB, with a false alarm when $|\Delta\hat L|\geq3$ dB. We first compute RMSE separately for each traffic--initialization pair and then average; false-alarm rates pool stable-sample counts.

\subsection{Vehicle Influence and Map Reconstruction}

A separate physical-characterization pool uses four previously generated traffic sequences and 768 future query area--time samples. Motion relative to the initial vehicle configuration produces changes of at least 3 and 6 dB in 12.24\% and 5.99\% of samples, with a maximum of 13.13 dB. Against a matched empty scene, the corresponding fractions are 17.84\% and 4.30\%. These localized effects (Fig.~\ref{fig:impact}) show that a vehicle can substantially alter reception along influential segments despite a large static background.

Table~\ref{tab:main} shows that OA-GS achieves map/strong-event RMSEs of 1.272/3.563 dB, improving on GS-CG by 1.232/2.402 dB under this protocol. Its event margin over LOCAL is smaller than its map margin. GS-CG retains the lowest stable-region false-alarm rate: 0.132\%, compared with 1.186\% for OA-GS.

\subsection{Contribution of the Shared Area Correction}
\begin{figure}[t]
\centering
\includegraphics[width=\columnwidth]{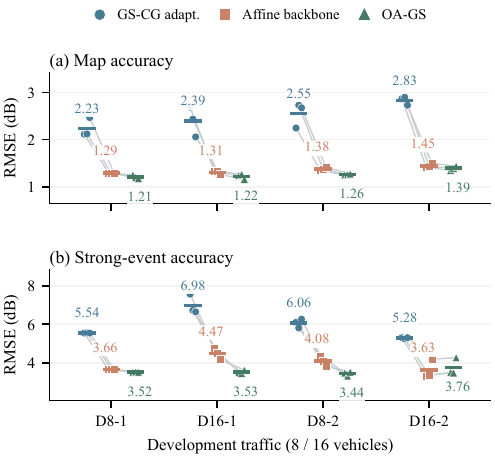}
\caption{Matched development results. Thin lines pair initializations; thick ticks and numbers show three-initialization means. OA-GS improves on its affine backbone in all 12 map comparisons and nine strong-event comparisons.}
\label{fig:paired}
\end{figure}
The same affine backbone without area correction achieves 1.355-dB map and 3.962-dB strong-event RMSE. Adding the shared correction lowers these errors by 0.083 and 0.399 dB. Figure~\ref{fig:paired} shows map improvement in all 12 pairs and event improvement in nine; all three initialization means improve. One 16-vehicle sequence has a 0.132-dB event regression. Stable-region false alarms increase from 0.461\% to 1.186\%. These matched results isolate the correction's contribution from the full-system baseline gap.

On a matched subset of two development sequences and three initializations, correction reduces strong-event RMSE by 0.483 dB under 0.8-s delay and 0.484 dB under doubled position bias/jitter. Increased position noise still degrades OA-GS.

%% file: tables/main_results.tex
\begin{tabular}{@{}lrrrr@{}}
\toprule
Method & Initial RF & Map & Strong & Quiet\\
 & & \multicolumn{2}{c}{RMSE (dB)$\downarrow$} & FPR (\%)$\downarrow$\\
\midrule
GS-CG adapt. & 8 & 2.503 & 5.965 & \textbf{0.13} \\
MLP-GS adapt. & 8 & 3.478 & 6.442 & 3.43 \\
LOCAL & 0 & 1.541 & 3.623 & 3.10 \\
Affine backbone & 0 & 1.355 & 3.962 & 0.46 \\
\textbf{OA-GS} & 0 & \textbf{1.272} & \textbf{3.563} & 1.19 \\
\bottomrule
\end{tabular}